\documentstyle{res}
\input epsf

\begin{document}
\def\smalltext#1{
\noindent{\small #1}
\baselineskip=14pt 
}
\def\spose#1{\hbox to 0pt{#1\hss}}
\def\simlt{\mathrel{\spose{\lower 3pt\hbox{$\mathchar"218$}}
     \raise 2.0pt\hbox{$\mathchar"13C$}}}
\def\simgt{\mathrel{\spose{\lower 3pt\hbox{$\mathchar"218$}}
     \raise 2.0pt\hbox{$\mathchar"13E$}}}
%\simpropto produces \propto with twiddle underneath
\def\simpropto{\mathrel{\spose{\lower 3pt\hbox{$\mathchar"218$}}
     \raise 2.0pt\hbox{$\propto$}}}

% redefine the labels 

\def\thesection {\arabic{section}}
\def\thefigure{\arabic{figure}}

% redefine thebibliography
     
\def\thebibliography#1{\section*{References}\baselineskip 13pt \parskip 0pt
\list
 {\arabic{enumi}.}{\settowidth\labelwidth{[#1]}\leftmargin\labelwidth
 \advance\leftmargin\labelsep
 \usecounter{enumi}}
 \def\newblock{\hskip .11em plus .33em minus -.07em}
 \sloppy
 \sfcode`\.=1000\relax 
}
\let\endthebibliography=\endlist

\title{%
CMB Anisotropies: A Decadal Survey}

\author{Wayne Hu \\
{\it    Institute for Advanced Study, Princeton, NJ, 08540, USA,
whu@ias.edu}
}

\maketitle
\begin{quotation}
    \footnotesize
    \noindent  
    Said the disciple, ``After I heard your words, one year
    and I ran wild, two years and I was tame, three years
    and positions interchanged, four years and things settled
    down, five years and things came to me, nine years
    and I had the great secret.''
    
    \begin{center}
    --Chuang-tzu
    \end{center}
    
\end{quotation}
    
\section*{Abstract}
\begin{quotation}
\footnotesize\noindent
We review the theoretical implications of the past decade of
CMB anisotropy measurements, which culminated in the recent
detection of the first feature in the power spectrum,
and discuss the tests available to the next decade of experiments.  
The current data already suggest that
density perturbations originated in an inflationary epoch, 
the universe is spatially flat,
and baryonic dark matter is required.  We discuss the underlying
assumptions of these claims and outline the tests required 
to ensure they are robust.  The most critical test - 
the presence of a second feature at the predicted location 
- should soon be available.  Further
in the future, 
secondary anisotropies and polarization should open new windows to 
the early and low(er) redshift universe.
\end{quotation}

\section{Introduction}

The 1990's will be remembered as a decade of discovery for
cosmic microwave background (CMB) anisotropies.  The launch
of the COBE satellite ushered in the decade in 1990 and lead 
to the first detection of CMB anisotropies
at $>10^{\circ}$ scales \cite{COBE92}.  Through the
decade, a combination of higher resolution experiments made the
case for a rise in the anisotropy level on 
degree scales and a subsequent fall at arcminute scales \cite{ScoSilWhi95}. 
The final year saw experiments, notably Toco and Boomerang,
with sufficient angular resolution
and sky coverage to localize a sharp peak in the anisotropy
spectrum at approximately $0^{\circ}\!.5$ \cite{TegZal00}.  In this review,
we discuss the theoretical implications of these results 
and provide a roadmap for critical tests and uses of
CMB anisotropies in the coming decade.

\section{Once and Future Power Spectrum}

The tiny $10^{-5}$ variations in the temperature of the CMB
across the sky are observed to be consistent with Gaussian
random fluctuations, at least on the COBE scales ($>10^{\circ}$),
as expected in the simplest theories of their inflationary
origin.  Assuming Gaussianity, the fluctuations can be fully
characterized by their angular power spectrum\footnote{$\!\!$Conventions for 
relating multipole number to angular
scale include: $\theta_{\ell} \approx 2\pi/\ell$, $\pi/\ell$ or
$100^{\circ}/\ell$.  
To the extent that these conventions
differ, none of them are correct; we hereafter refer to power
spectrum features by multipole number, which has a precise meaning.}
\begin{equation}
    T(\hat{\bf n}) = \sum_{\ell m} a_{\ell m} Y_{\ell m}(\hat{\bf n})\,,
   \qquad \left< a_{\ell m}^{*} a_{\ell 'm'} \right> =
   \delta_{\ell \ell'}\delta_{m m'} C_{\ell}\,.
\end{equation}
We will often use the shorthand $(\Delta T)^{2} = 
\ell (\ell +1)C_{\ell}/2\pi$ which represents the power
per logarithmic interval in $\ell$.  

\begin{figure}[t]
\begin{center}
\epsfxsize=\textwidth\epsffile{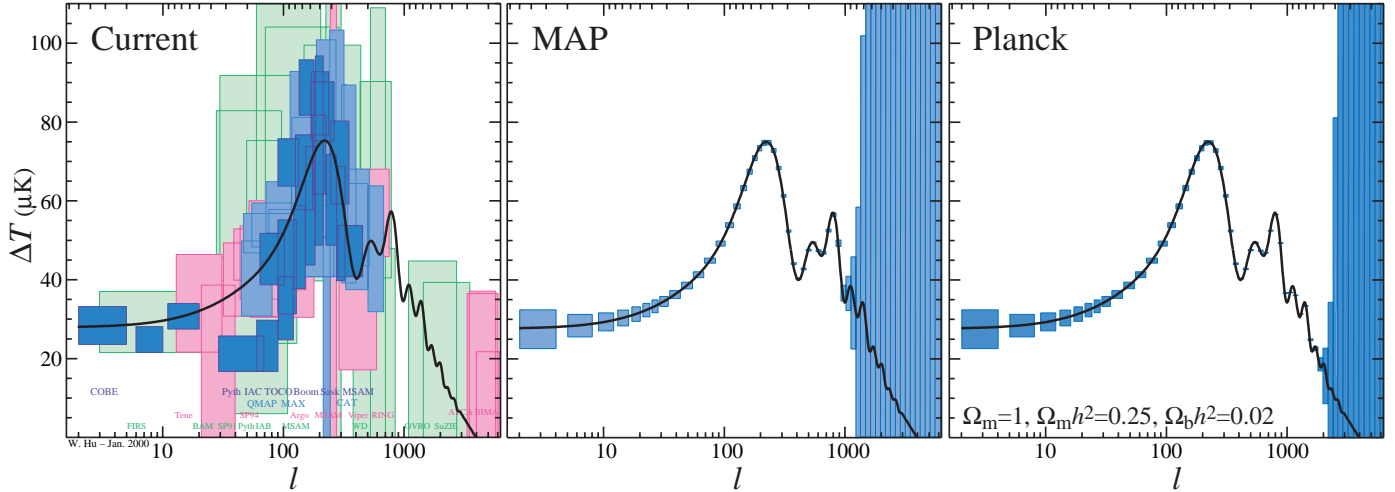}
\end{center}
\vspace{-0.8truecm}
\caption{$\!\!$: Power Spectrum}
%\smalltext{Power Spectrum. Error boxes represent $1\sigma$ in power
%and the FWHM of the experimental windows for the current 
%data and band power range for the MAP and Planck satellites.}
\label{fig:power}
\end{figure}

Fig.~\ref{fig:power} (left, $1\sigma$ errors $\times$ window FWHM) 
shows 
the measurements the power spectrum to date
(see \cite{TegZal00} for a complete list of references).  
The data indicate a rather sharp peak in the spectrum
at $\ell \sim 200$ with a significant decline at $\ell \simgt 1000$.
This peak has profound implications for
cosmology.  The primary claims in decreasing order of
confidence and increasing need of verification from precision
measurements (e.g. from the MAP and Planck satellite 
Fig.~\ref{fig:power} center, right) are 

\begin{itemize}
    
    \footnotesize
\item{} Early universe. The simplest inflationary cold dark matter (CDM) cosmologies
have correctly predicted the location and morphology of the 
first peak in the CMB; conversely,
all competing {\it ab initio} theories have failed, essentially due
to causality. {\it 
Confirm its acoustic nature with the second peak. 
Use polarization as a sharp test of causality.}

\item{} Geometry. The universe is flat.  {\it Lower} limits on the
total density ($\Omega_{tot}\equiv 
\sum \Omega_i \simgt 0.6$ \cite{TegZal00}) are already robust, 
unless recombination is substantially
delayed or $h \gg 1$.  {\it Calibrate
the ``standard rulers'' (acoustic scale and damping scale)
in this distance measure through the higher peaks}. 

\item{} Baryons.  At least as much baryonic
dark matter  as indicated by big 
bang
nucleosynthesis (BBN) is required ($\Omega_{b} h^{2}
\simgt 0.01$ \cite{TegZal00}). {\it Confirm with relative heights of
the peaks, especially the third peak.}

\item{} Reionization. The Thomson optical depth is low -- how low
depends on the range of models considered.  {\it The optical depth,
assuming it is low, 
will only be accurately measured by CMB polarization at large angles.}

\item{} Dark energy. The matter density is low and combined with 
flatness, this indicates
a missing energy component, possibly the cosmological constant. 
Currently the $95\%$ CL includes $\Omega_{m}=1$ but the maximum 
likelihood model including BBN and $h$ 
constraints has $\Omega_{m}\approx 0.3$ \cite{TegZal00}.
{\it Measure $\Omega_{m}h^{2}$ from the
first three peaks.} 

\end{itemize}
The early universe and geometry tests basically rely on the
position of the first peak and hence are more robust than the 
later ones which rely mainly on interpreting its amplitude.

Moreover, all claims are based on interpreting the
peak at $\ell \sim 200$ as the first in a series of {\it acoustic
peaks}.   Based on the sharpness of the feature, this interpretation
is now reasonably, but not completely secure. 
The detection of a second peak in the spectrum is critical 
since it will
provide essentially incontrovertible evidence that this 
interpretation is correct (or wrong!).  
Once this is achieved
and the peaks pass the morphological tests described below, the CMB
will become the premier laboratory for precision
cosmology, as many studies have shown
\cite{Junetal96}.
These expectations also rely on the fact that $C_\ell$
can ultimately be measured to 
\begin{equation}
{\Delta C_{\ell} \over C_{\ell}} = \sqrt{2 \over (2\ell +1)f_{\rm sky}}\,,
\label{eqn:sample}
\end{equation}
based on
Gaussian sample variance on the 
$(2\ell+1)f_{\rm sky}$ independent modes of a given $\ell$,
from a fraction of sky $f_{\rm sky}$.
The rest of this review will make the theoretical
case for the above statements.

\begin{figure}[t]
\begin{center}
\epsfxsize=\textwidth\epsffile{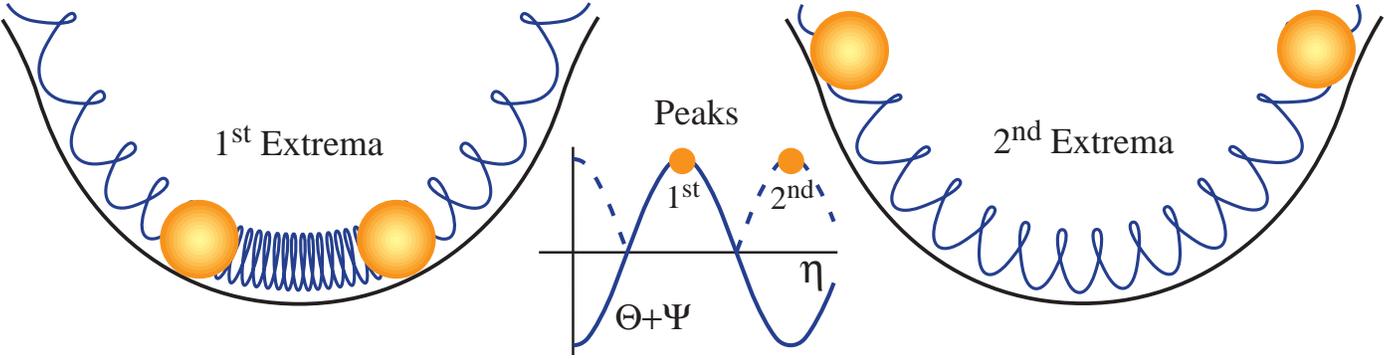}
\end{center}
\vspace{-0.8truecm}
\caption{$\!\!$: Harmonic Acoustic Peaks.}
{\footnotesize }
\label{fig:harmonics}
\end{figure}

\section{Sound Physics}

The theory underlying the predictions of CMB anisotropies has
essentially been in place since the 1970's \cite{Early} and is
based on extraordinarily simple fluid mechanics and gravity 
\cite{Fluid,Sel94}.
Simplicity is ensured by the smallness of the fluctuations 
themselves: the observed amplitude of $\Delta T/T \sim 10^{-5}$
guarantees that the equations of motion can be linearized. 

The fluid nature of the problem follows from simple thermal
arguments. The cooling of CMB photons due to the cosmological expansion
implies that before $z_*\approx 1000$,
when the CMB temperature is $T > 3000$K, the photons are hot enough to
ionize hydrogen.  During this epoch, the electrons glue the baryons
to the photons by Compton scattering and electromagnetic interactions.
The dynamics that result involve a single photon-baryon fluid.

Gravity attracts and compresses the fluid into the potential wells
that later seed large-scale structure.
Photon pressure resists this compression and sets up
sound waves or acoustic oscillations in the fluid.
These sound waves are frozen into the CMB at recombination.
Regions that have reached their maximal compression by recombination
become hot spots on the sky; those that reach maximum rarefaction
become cold spots.

\section{Math}
\label{sec:oscillator}

Mathematically, the cast of characters are: for the photons,
the local temperature $\Theta=\Delta T/T$, bulk velocity or dipole
$v_{\gamma}$, and anisotropic stress or quadrupole $\pi_{\gamma}$;
for the baryons, the density perturbation $\delta_{b}$ and bulk 
velocity $v_{b}$; for gravity, the Newtonian potential $\Psi$ 
(time-time metric fluctuation) and the curvature fluctuation
$\Phi$ (space-space metric fluctuation $\approx -\Psi$).  
Covariant conservation of energy 
and momentum requires that the photons and baryons satisfy seperate continuity
equations \cite{Fluid}
\begin{eqnarray}
\dot \Theta = -{k \over 3} v_{\gamma} - \dot\Phi \, , \qquad
\dot \delta_b = -k v_b - 3\dot\Phi \, ,
\label{eqn:continuity}
\end{eqnarray}
and Euler equations
\begin{eqnarray}
\dot v_{\gamma} &=& k(\Theta + \Psi) - {k \over 6} [1+3(1-\Omega_{\rm 
tot}) {H_0^2 \over k^2}]
\pi_\gamma
        - \dot\tau (v_\gamma - v_b) \, , \nonumber\\
\dot v_b &= &- {\dot a \over a} v_b + k\Psi + \dot\tau(v_{\gamma} - v_b)/R
        \, ,
\label{eqn:Euler}
\end{eqnarray}
in wavenumber space. $\dot\tau = n_e\sigma_T a$ is the differential Thomson
optical depth,  
$R= (p_b + \rho_b)/(p_\gamma + \rho_\gamma) \approx
3\rho_b / 4 \rho_\gamma$ is the photon-baryon momentum density ratio, and 
overdots represent derivatives with respect to conformal
time $\eta=\int dt/a$.

The continuity equations represent particle number conservation.
For the baryons, $\rho_b \propto n_b$.
For the photons, $T \propto n_\gamma^{1/3}$, which explains the
$1/3$ in the velocity divergence term. 
The $\dot\Phi$ term represents
the ``metric stretching'' effect and appears because $\Phi$ represents
a spatially varying perturbation to the scale factor $a$ and
$n_{\gamma,b} \propto a^{-3}$ (see Fig.~\ref{fig:secondaries}, left).

The Euler equation has a similar interpretation. 
The expansion makes particle momenta
decay as $a^{-1}$.  The cosmological redshift of $T$ accounts for
this effect in the photons.
For the baryons, it becomes the expansion drag on 
$v_b$ ($\dot a/a$ term).  Potential gradients $k\Psi$ generate 
potential flow.  For the photons, stress gradients in the fluid, both
isotropic ($k\delta p_\gamma /(p_\gamma + \rho_\gamma) = k
\Theta$) and anisotropic ($k\pi_\gamma$) counter infall.  
Compton scattering
exchanges momentum between the two fluids
($\dot \tau$ terms). 

If scattering $(\dot\tau^{-1})$
is rapid compared with the light travel time
across the perturbation $(k^{-1})$,  the photon-baryon system
behaves as a perfect fluid.  To lowest order in $k/\dot\tau$,
eqns.~(\ref{eqn:continuity}) and (\ref{eqn:Euler})
become
\begin{equation}
(m_{\rm eff} \dot\Theta)\dot{\vphantom{A}} + {k^2 \over 3}\Theta
= -{k^2 \over 3} m_{\rm eff}\Psi - (m_{\rm eff}\dot\Phi)
\dot{\vphantom{A}} \,,
\label{eqn:oscillator}
\end{equation}
%\begin{eqnarray}
%\dot \Theta &=& - {k \over 3} \Theta_1 - \dot\Phi  \, ,\nonumber\\
%\dot v_\gamma &=& -{R \over 1 + R} {\dot a \over a} v_\gamma +
%        {1 \over 1 + R} k\Theta_0 + k\Psi.
%\end{eqnarray}
where the effective mass is $m_{\rm eff} = 1+R$ or alternatively
$c_s^2 = \dot p/\dot\rho = 1/3m_{\rm eff}$.
Scattering isotropizes the distribution in the
electron rest frame $v_\gamma=v_b$ 
and eliminates anisotropic stress ($\pi_\gamma = {\cal O}(k/\dot\tau) v_\gamma$). 

Equation (\ref{eqn:oscillator}) is the fundamental relation for 
acoustic oscillations; it 
reads: the change in the momentum of the photon-baryon
fluid is determined by a competition between the
pressure restoring and gravitational driving forces.
Given the initial conditions and gravitational potentials,
it predicts the phenomenology of the acoustic peaks.

\begin{figure}[t]
\begin{center}
\epsfxsize=\textwidth\epsffile{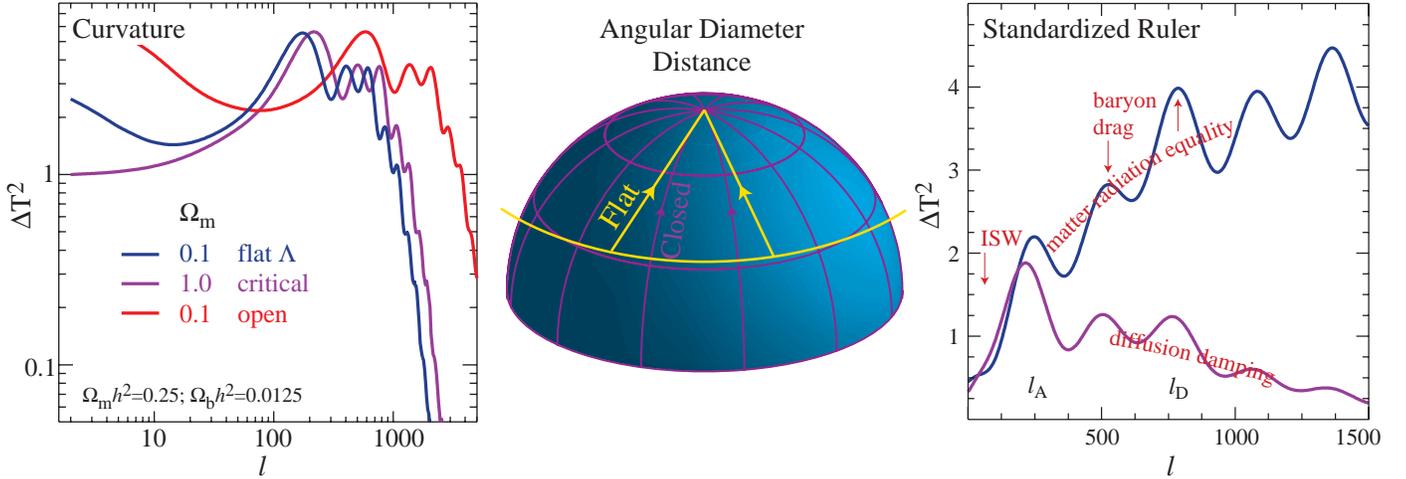}
\end{center}
\vspace{-0.8truecm}
\caption{$\!\!$: Geometry.}
{\footnotesize}
\label{fig:geometry}
\end{figure}

\section{Early Universe}
\label{sec:early}

The simplest inflationary models are essentially unique in their
phenomenological predictions.  They possess a spectrum
of curvature (potential) fluctuations that extends
{\it outside} the apparent horizon in the post-inflationary epoch. 
These perturbations remain constant while the fluctuation is outside
the horizon except for a small change at matter-radiation equality.  
Neglecting this and baryon inertia $(m_{\rm eff}=1)$ for the moment,
the oscillator equation (\ref{eqn:oscillator}) 
has the simple solution
\begin{equation}
[\Theta + \Psi](\eta_*) = [\Theta + \Psi](0) \cos(k s)\,,
\qquad 
v_\gamma = \sqrt{3} [\Theta + \Psi](0) \sin(k s)\,,
\end{equation}
where $s=\int_0^{\eta_*} c_s d\eta$ is the {\it sound horizon}
at $\eta_*$.  An initial temperature perturbation $\Theta(0)$ 
exists since the gravitational
potential $\Psi$ is a time-time perturbation to the metric.  Because
of the redshift with the scale factor $a \propto t^{2/3(1+p/\rho)}$,
a temporal shift produces a temperature perturbation of
$\Theta = -2 \Psi/3(1+p/\rho)$ or $-\Psi/2$ in the radiation dominated
era. We call $\Theta+\Psi$ the effective temperature since it
also accounts for the redshift a photon experiences when climbing out
of a potential well \cite{SacWol67}.  The matter radiation transition simply 
makes $\Theta +\Psi  = \Psi /3$.

There are two important aspects of this result.  First, inflation sets
the {\it temporal} phase of all wavemodes by starting them all at 
the initial epoch.  Wavenumbers which hit their extrema at recombination
are given by $k_m = m\pi/s$ and these mark the peaks of coherent oscillation
in the power spectrum.  
Second, the first peak at $k=\pi/s$ represents a
compression of the fluid in the gravitational potential well ($\Psi<0$,
see Fig.~\ref{fig:harmonics}).

Without inflation to push perturbations
superluminally outside the horizon, they 
must be generated by the causal motion of matter.  One might think
any anisotopies above the horizon scale 
at recombination projected on the sky (e.g. COBE) implies 
inflation.  
However these could instead be generated after
recombination through gravitational redshifts (\S \ref{sec:beyond}).  To test inflation,
one needs to isolate
a particular epoch in time.  The acoustic peaks provide one
such opportunity; we shall see later that polarization provides 
another.

If the fluctuations were generated
by non-linear dynamics well inside the horizon,
e.g. by a cosmic string network, the temporal coherence,
and hence the peak structures,
would be lost due to random forcing of
the oscillators \cite{AlbCouFerMag96}.  
Causal generation itself does not guarantee incoherence. 
Coherence requires that there is one special epoch for all modes that
synchs up their oscillations. One common event can 
causally achieve this: horizon crossing when $k\eta=1$.  
For example, textures unwind at horizon
crossing and maintain some coherence in their acoustic oscillations. 
However it is very difficult to
place the first compressional peak at as large a scale as 
$k_1 = \pi/s$ since the photons tend to first cool down due to
metric stretching from $\Phi$ 
as gravitational potentials grow, thus inhibiting the
compressional heating \cite{HuWhi96}.   
The only known mechanism for doing so 
is to reverse the sign of gravity: to make gravitational
potential wells in underdense regions so that $\Phi \sim \Psi$ 
\cite{Tur96}.
In principle, this can be arranged by a special choice of anisotropic
stresses but there is no known form of matter that obeys the required
relations.  On the other hand, {\it inflationary} curvature 
(adiabatic) 
and isocurvature (stress) fluctuations existing outside the horizon
can be interconverted with physically realizable stress histories \cite{Hu99}.

In summary, verification of an inflationary series of acoustic peaks
with locations in an approximate ratio of $\ell_1:\ell_2:\ell_3\ldots=
1:2:3\ldots$ would represent
a strong test of the inflationary origin of the perturbations and 
a somewhat weaker test of their initially adiabatic nature.

\begin{figure}[t]
\begin{center}
\epsfxsize=\textwidth\epsffile{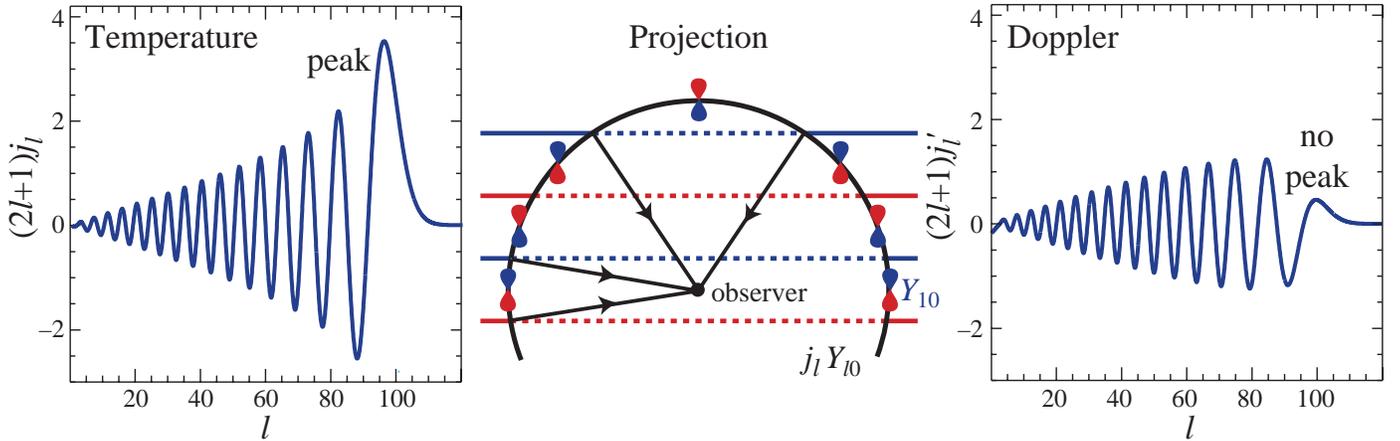}
\end{center}
\vspace{-0.8truecm}
\caption{$\!\!$: Acoustic not Doppler peaks.}
\label{fig:project}
\footnotesize
\end{figure}

\section{Geometry}

The physical scale of the features is related to the distance $s$ that
sound can travel by recombination.  Specifically, one expects features
in the spatial power spectrum of the photon temperature and dipole
at $k> k_A = \pi/s$.  Each mode is then projected
on the sky in spherical coordinates 
$\exp(i{\bf k}\cdot{\bf x}) \propto j_\ell(kd) Y_{\ell 0}$, where
$d = \eta_0-\eta_*$, and summed in quadrature to form the final anisotropy,
\begin{equation}
C_\ell \approx {2 \over \pi} \int {dk \over k} k^3
\left[ (\Theta + \Psi) j_\ell(kd) + v_\gamma j_\ell'(kd)\right]^2\,.
\label{eqn:clproject}
\end{equation}
This approximation ignores the finite duration of recombination but suffices
for a qualitative understanding of the spectrum.  We 
have also temporarily assumed that the universe 
is flat $\Omega_{\rm tot}=1$.

The $v_\gamma$ term represents the Doppler effect from the 
motion of the
fluid along the line of sight.  It has an intrinsic
dipole angular dependence at last scattering $Y_{10}$ in addition to
the ``orbital'' angular dependence $Y_{\ell 0}$.  Addition of
angular momentum implies a coupling of $j_{\ell \pm 1}$ that can be
rewritten as $j_\ell'$.  

As a consequence of eqn.~(\ref{eqn:clproject}), features in the
spatial power spectrum of the effective temperature at recombination
become features in the angular power spectrum whereas those of
the bulk velocity do not (see Fig.~\ref{fig:project} $kd=100$)
\cite{Fluid}.  
A plane wave temperature perturbation contributes a range of anisotropies
corresponds to viewing angles perpendicular ($\ell \approx kd$) all the
way to parallel ($\ell \rightarrow 0$) to the wavevector ${\bf k}$
(see Fig.~\ref{fig:project}, lobes).  
The result is a sharp maximum around $\ell = kd$ as expected from 
naively converting physical to angular scale.   However for
the Doppler effect from potential flows, velocities are directed parallel to 
${\bf k}$, so that the peak at $\ell = kd$ 
is eliminated.  Although the Doppler effect contributes
significantly to the overall anisotropy, the peak structure traces
the temperature fluctuations. 

In a spatially curved universe, one replaces the spherical Bessel
functions in eqn.~(\ref{eqn:clproject}) with the ultraspherical 
Bessel functions and these peak at $\ell \approx kD$ where $D$
is the {\it comoving angular diameter distance}
to recombination. 
Consider first 
a closed universe with radius of curvature 
${\cal R} =  H_{0}|\Omega_{\rm tot}-1|^{1/2}$.
Suppressing one spatial coordinate yields
a 2-sphere geometry with the observer situated at the
pole (see Fig.~\ref{fig:geometry}).  Light travels on 
lines of longitude.
A physical scale $\lambda$ at fixed latitude given by
the polar angle $\theta$ subtends an angle
$\alpha = \lambda/{\cal R}\sin\theta$.
For $\alpha \ll 1$,
a Euclidean analysis would infer a 
distance $D={\cal R}\sin\theta$, even though
the {\it coordinate distance} along the arc is
$d = \theta {\cal R}$; thus
\begin{equation}
D = {\cal R} \sin( d / {\cal R})\,, \qquad (\Omega_{\rm tot}>1)\,.
\end{equation}
For open universes, simply replace $\sin$ with $\sinh$.
A given physical scale subtends a larger (smaller) angle in
a closed (open) universe than a flat universe.

We thus expect CMB features at the 
characteristic scale \cite{HuSug95}
\begin{eqnarray}
    \ell_{A} &=& \pi D/s 
	\approx 172 \Omega_{\rm tot}^{-1/2} 
	[1+\ln(1-\Omega_\Lambda)^{0.085}]
	f(\Omega_m h^2,\Omega_b h^2)\,,\\
\label{eqn:ellA}
f &=& \left({z_* \over 10^3}\right)^{1/2} \left( {1 \over \sqrt{R_*}} \ln 
	{\sqrt{1+R_*} +  \sqrt{R_* +\epsilon R_*} 
	\over  1 + \sqrt{\epsilon R_*}} \right)^{-1}  \,,
\label{eqn:correction}
\end{eqnarray}
where $\epsilon \equiv a_{eq}/a_* = 0.042 (\Omega_m h^2)^{-1} (z_*/10^3)$
and $R_* = 30\Omega_b h^2 (z_*/10^3)$; see \cite{HuWhi97} for
$z_*(\Omega_m h^2,\Omega_b h^2)$. 

The main scaling of $\ell_{A}$
is with $\Omega_{\rm tot}^{-1/2}$ \cite{KamSpeSug94},
but finite $\Omega_{\Lambda}$
causes it to decrease.
This covariance is referred to in the literature as the {\it angular
diameter distance $(D)$ degeneracy}.  
The quantity in 
parentheses in eqn.~(\ref{eqn:correction}) goes to unity 
as $\epsilon,R_{*}\rightarrow 0$. The leading order 
correction ($1+\epsilon^{1/2}$) makes the $\Omega_m h^2$ dependence
important in any reasonable cosmology. 
The other correction ($1+R_*/6$) is small for reasonable 
baryon densities.

For simple inflationary models, the peaks reside at $\ell_m \approx m \ell_A$.
More generally,  $\ell_1 \ge \ell_A$ (see \S \ref{sec:early}). 
The detection of the first peak then
puts a reasonably robust lower limit on $\Omega_{\rm tot}$.
The key assumptions are that we can attribute the feature
to acoustic oscillations, bound the redshift of recombination from below and bound the sound
horizon from above.  The last assumption amounts to 
having an upper limit on $\Omega_{m} h^2$ (or $h$). 
The $D$ degeneracy is tamed since $\Omega_\Lambda$ is automatically bounded 
from above for the $\Omega_{\rm tot}$ of interest by requiring
$\Omega_m >0$.
Converting lower limits on $\Omega_{\rm tot}$ into precise measurements
requires independent measurements of
$\Omega_m h^2$ and $\Omega_b h^2$, which calibrate the standard rulers at recombination \cite{Fluid}, and
$\Omega_\Lambda$, $\Omega_m$ or $h$ to break the $D$ degeneracy.

\begin{figure}[t]
\begin{center}
\epsfxsize=\textwidth\epsffile{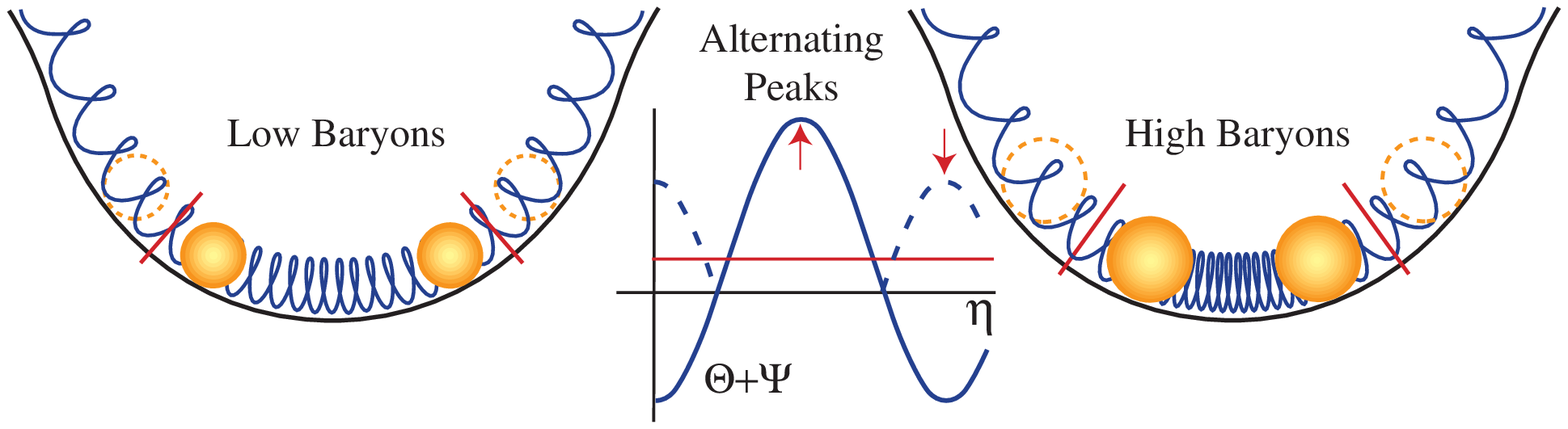}
\end{center}
\vspace{-0.8truecm}
\caption{$\!\!$: Baryons.}
\label{fig:baryons}
\footnotesize
\end{figure}

\section{Baryons}

Baryons add inertia to the fluid.
Consider first the case of
$m_{\rm eff} = 1+R=$ const. [see
eqn.~(\ref{eqn:oscillator})]
\begin{equation}
[\Theta + \Psi](\eta_*) = [\Theta(0) + (1+R)\Psi(0)]
	\cos(ks) - R\Psi\,,
\end{equation} 
where $s=\eta_*/\sqrt{3(1+R)}$.  There
are three effects of raising the baryon content:
an amplitude increase, a zero-point shift, and a frequency
decrease \cite{Fluid}.
Baryons drag the fluid deeper into the potential wells
(see Fig.~\ref{fig:baryons}).  For the fixed initial conditions,
the resulting shift in the zero point also implies a
larger amplitude.  Since it is the power spectrum that is observed,
the result of squaring implies that all compressional peaks are 
enhanced by the baryons and the rarefaction peaks suppressed.  This 
is the clearest signature of the baryons and also provides a 
means for testing the compressional nature of the first peak predicted by inflation.    
The fact that $R \propto a$ due to the redshifting of the photons
simply means that the oscillator actually has time dependent mass.  
The adiabatic invariant ($E/\omega$) implies an 
amplitude reduction as $(1+R)^{-1/4}$.

Baryons also affect the fluid through dissipational processes
\cite{Sil68}.
The random walk of the photons through the baryons
damps the acoustic oscillation exponentially below  
the diffusion scale $k_D$, roughly the geometric 
mean of the mean free path and the horizon scale.
Microphysically, the dissipation comes from viscosity
$\pi_\gamma$ in eqn.~(\ref{eqn:Euler}) and heat conduction $v_\gamma-v_b$.
Before recombination it can be included by keeping terms of order
$k/\dot\tau$ in the equations.  At recombination, the 
mean free path increases and brings the diffusion scale to \cite{HuWhi97}
\begin{eqnarray} 
k_D &\approx& a_1(\Omega_{m}h^{2})\,
(\Omega_b h^2)^{0.291} [1+a_2(\Omega_{m}h^{2})\,(\Omega_b h^2)^{1.8}]^{-1/5} 
	{\rm Mpc}^{-1}\,, 
\end{eqnarray}	
$a_1(x) = 0.0396x^{-0.248}(1+ 13.6x^{0.638})$,
$a_2(x) = 1480x^{-0.0606} (1+ 10.2x^{0.553})^{-1}$.
The main effects can be easily understood: increasing $\Omega_m h^2$
decreases the horizon at last scattering 
and hence the diffusion length.
At low $\Omega_b h^2$, increasing the baryon content decreases
the mean free path while at high $\Omega_b h^2$, it delays recombination
and increases the diffusion length.  

Damping introduces another length scale for the curvature test, $l_D = 
k_D D$; alternately 
$l_D/l_A = k_D/k_A = f(\Omega_m h^2,\Omega_b h^2)$ is
independent of $D$ and can measure this 
combination of parameters.

\begin{figure}[t]
\begin{center}
\epsfxsize=\textwidth\epsffile{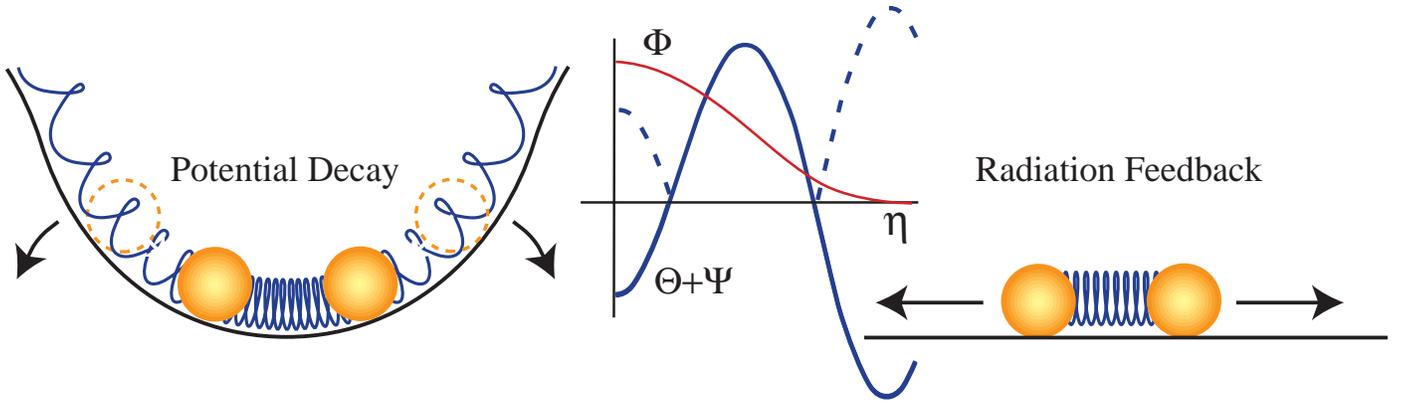}
\end{center}
\vspace{-0.8truecm}
\caption{$\!\!$: Matter-radiation ratio.}
\label{fig:drive}
\footnotesize
\end{figure}

\section{Matter/Radiation}
\label{sec:matterradiation}

We have hitherto been considering the gravitational force 
on the 
oscillators as constant in time.  This can only be true for 
{\it growing} density fluctuations.
The Poisson equation says that $\Phi \propto a^2 \rho \delta$,
and the density redshifts with the expansion as $\rho \propto 
a^{-3(1+p/\rho)}$.  In the radiation era, density perturbations must
grow as $a^2$ for constant potentials, as they do in the comoving
gauge when pressure gradients can be neglected.  Once
the pressure gradients have turned infall into acoustic oscillations,
the potential must decay.   This decay actually drives the
oscillations since the fluid is left maximally compressed with
no gravitational potential to fight as it turns around 
(see Fig.~\ref{fig:drive}) \cite{Fluid}. 
The net effect is doubled by the metric stretching effect from $\Phi$,
leading to fluctuations with amplitude 
$2\Psi(0) - [\Theta +\Psi](0) = {3 \over 2}\Psi(0)$.

When the universe becomes matter-dominated the gravitational potential
no longer reflects photon density perturbations. As discussed in \S 
\ref{sec:early}, $\Theta +\Psi =
{\Psi/3} = 3\Psi(0)/10$ here, so that across the horizon scale at
matter radiation equality the acoustic amplitude increases by a factor of 5.  

This effect mainly measures the matter-to-radiation ratio.  Density perturbations
in any form of radiation will stop growing around horizon crossing
and lead to this effect.  For the neutrinos, the only difference is 
that
anisotropic stress from their quadrupole anisotropies 
also slightly affects the cessation
of growth.  
One can only turn this into a measure of $\Omega_m h^2$
by assuming that the radiation density is known through the CMB
temperature and the number of neutrino species otherwise we are
faced with a {\it matter-radiation}
degeneracy. For example,  determining
both $\Omega_m h^2$ and the number of neutrino species from the CMB
alone will be difficult.

Precise measurements of $\Omega_m h^2$ when combined with the
angular diameter distance would constrain the universe to live on 
a line in the
classical cosmological parameter space ($\Omega_m$,$\Omega_\Lambda$,$h$).
{\it Any} external (non-degenerate) measurement in this space 
$(\Omega_m, h,$ acceleration,$\ldots$) and allows the three parameters 
to be determined independently.  This fortunate situation
has been dubbed ``cosmic complementarity'' and currently shows
``cosmic concordance'' around a $\Lambda$CDM model. 
More importantly, the combination of several checks
creates sharp consistency checks that may even show our universe
to live outside this space, for example if the missing energy 
is not $\Lambda$ but some dynamical ``quintessence'' field.

\section{And Beyond}
\label{sec:beyond}

The primary anisotropies from the recombination epoch contain
only a small fraction of the cosmological information 
latent in the CMB.
Let us conclude this
survey with topics of future study: 
secondary anisotropies and polarization.  
Both are expected to be at the $\simlt 10^{-6}$ ($\mu$K)
level and will require high sensitivity experiments with
wide-frequency coverage to reject galactic and extragalactic 
foregrounds of comparable amplitude.

\begin{figure}[t]
\begin{center}
\epsfxsize=\textwidth\epsffile{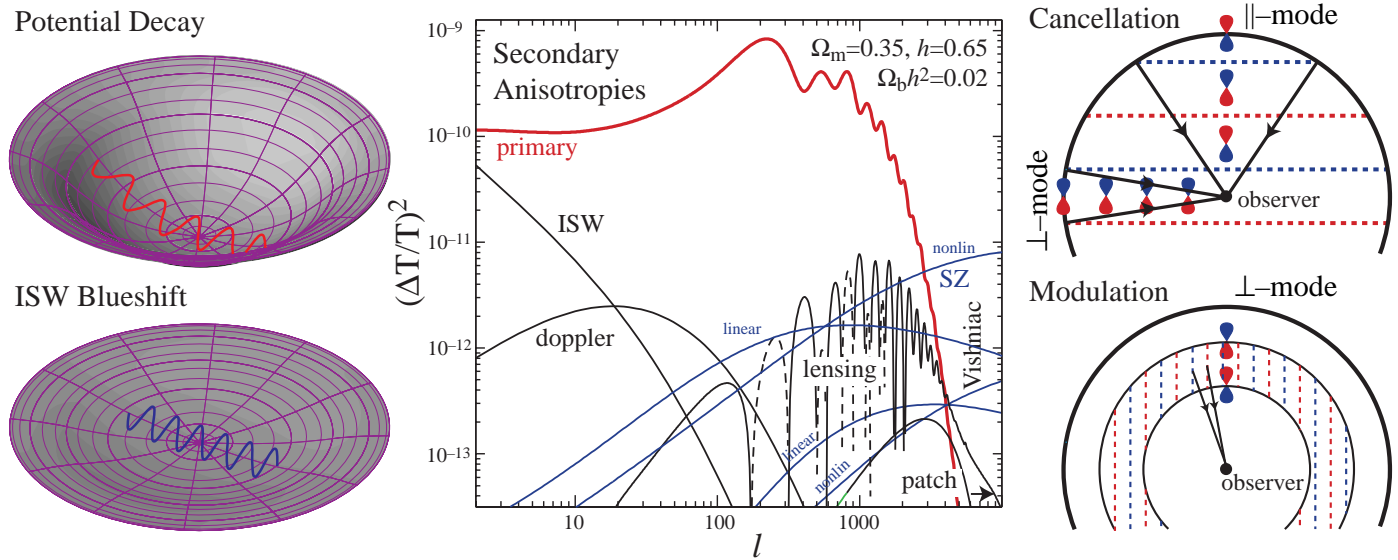}
\end{center}
\vspace{-0.8truecm}
\caption{$\!\!$: Secondary anisotropies.}
\label{fig:secondaries}
\footnotesize
\end{figure}

\vskip 0.2truecm
\noindent
{\bf Secondary Anisotropies:}
These are generated as photons travel
through the large-scale structure between us and recombination.
They arise from two sources: gravity and 
scattering during reionization.  It is currently believed that
the universe reionized at $5 \le z \simlt 15$ leading to $\tau_{\rm rei} \sim 0.01-0.1$.

Gravitational redshifts can change the temperature along the
line of sight.  Density perturbations cease to grow once
either the cosmological constant or curvature dominates the
expansion.  As discussed in \S \ref{sec:matterradiation}, 
the gravitational potentials must then decay.  Decay of potential
well both removes the gravitational redshift and heats the photons
by ``metric stretching" leading to an effect that is $2\Delta\Phi$
(see Fig.~\ref{fig:secondaries}).  The opposite effect occurs
in voids so that on small scales the anisotropies
are cancelled across crests and troughs of
modes parallel to the line-of-sight.
The effect from the decay is called the ISW effect \cite{SacWol67} and
from the non-linear growth of perturbations, the Rees-Sciama 
\cite{ResSci68} effect.

The gravitational potentials also
lens the CMB photons \cite{Lensing}.  Since lensing conserves surface 
brightness, it only affects
anisotropies and hence is second order.
The photons are deflected according to
the angular gradient of the potential integrated along the
line of sight. Again the cancellation of parallel modes
implies that large-scale potentials are mainly 
responsible for lensing and cause a long-wavelength modulation of
the sub-degree scale anisotropies.  The modulation is a power 
preserving smoothing of the power spectrum which reduces
the acoustic peaks to fill in the troughs.  Not until
the primary anisotropies disappear beneath the damping scale do
the cancelled potentials actually generate power in
the CMB.

The same principles apply for scattering effects -- with one twist.
The Doppler effect from large-scale potential flows, which
run parallel to the wavevector, contribute
nothing to the cancellation-surviving perpendicular modes
(see Fig.~\ref{fig:secondaries}).
Thus even though $v_{b}\tau \sim 10^{-4}-10^{-5}$, Doppler
contributions are at $10^{-6}$.  
The main effect of reionization is to suppress power in the
anisotropies as $e^{-2\tau}$ below the angle subtended by the
horizon at the scattering.  Unfortunately, given the sample
variance of the low-$\ell$ multipoles [see eqn.~(\ref{eqn:sample})], this effect is nearly
degenerate with the normalization and the current limits from
the first peak that $\tau_{\rm rei} \simlt 1$
will not be
improved by more than a factor of a few from the higher peaks.

Surviving the Doppler cancellation
are higher order effects due to optical depth modulation,
perpendicular to the line of sight, of
the Doppler shifts at small angular scales 
from linear density perturbations
(Vishniac effect \cite{Vis87}), 
non-linear structures (non-linear Vishniac
effect or kinetic SZ effect \cite{Hu00}) and patchy or inhomogeneous
reionization \cite{Patch}.  Another opacity-modulated signal is
the distortion from Compton upscattering by hot gas, the (thermal)
Sunyaev-Zel'dovich (SZ) effect \cite{SunZel80}, especially in clusters where it
is now routinely detected.

All of these secondary effects produce signals in the $\mu$K
regime. Developing methods to isolate them is currently an
active field of research and lies beyond the scope of
this review.  The main lines of inquiry are to explore
sub-arcminute scales where the primary anisotropies has fallen off,
the non-Gaussianity of the higher order effects \cite{NonGaus},
their frequency dependence to separate them from foregrounds and
the thermal SZ effect \cite{Fore}, their
cross correlation with other tracers of large-scale structure
\cite{Cross},
and finally their polarization.

\begin{figure}[t]
\begin{center}
\epsfxsize=\textwidth\epsffile{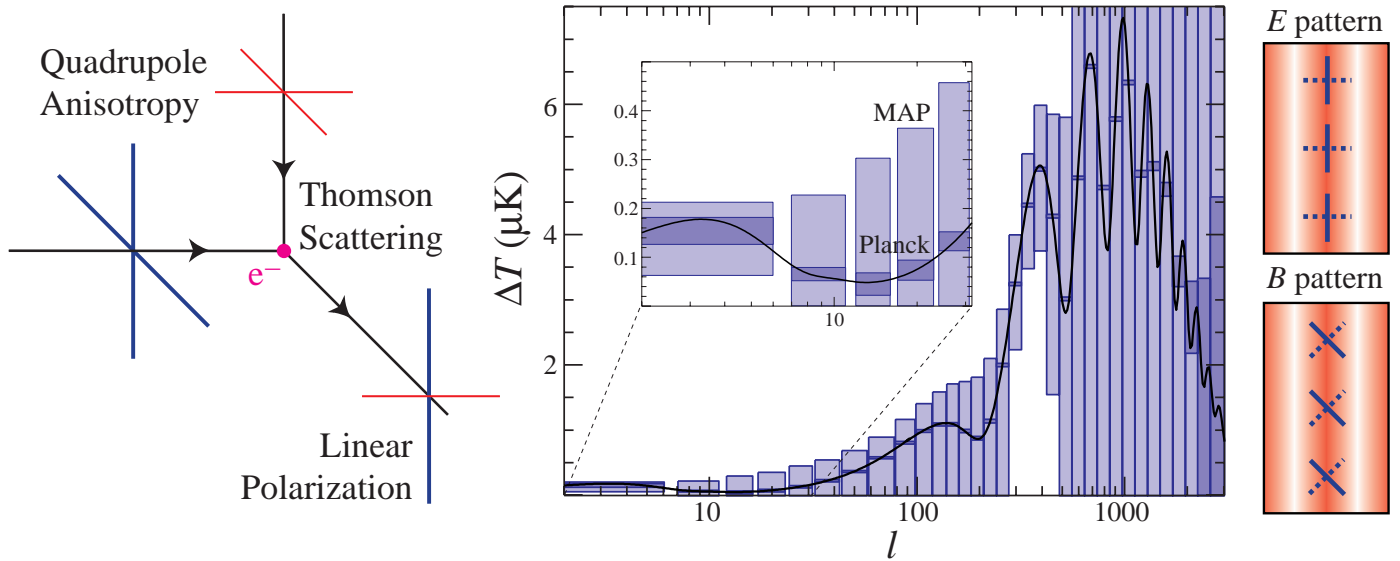}
\end{center}
\vspace{-0.8truecm}
\caption{$\!\!$: Polarization.}
\label{fig:pol}
\footnotesize
\end{figure}

\vskip 0.2truecm
\noindent
{\bf Polarization}:
Thomson scattering of quadrupole anisotropies generates
linear polarization in the CMB by passing only one component
of polarization of the incident radiation (see Fig.~\ref{fig:pol}).
The polarization amplitude, pattern, and correlation with the temperature
anisotropies themselves 
is thus encapsulated in the quadrupole 
anisotropies at the scattering. This information and the fact
that it is only generated by scattering are the useful properties 
of polarization.

Density perturbations generate quadrupole anisotropies as radiation
from
crests of a temperature perturbation flows into troughs.
Such anisotropies are azimuthally symmetric around the wavevector
($Y_{20}$ quadrupole).  They generate a distinct pattern where
the polarization is aligned or perpendicular to the wavevector
(``$E$'' pattern \cite{Pol}).

However polarization generation suffers
from a catch-22: the scattering which generates polarization also
suppresses its quadrupole source (see~\S \ref{sec:oscillator}).
They can only be generated once the perturbation becomes
optically thin.  Primary anisotropies are only substantially
polarized in the damping region where the finite duration of
last scattering allows viscous imperfections in the fluid, and
then only at the $\sim 10\%$ level ($\mu$K level, 
Fig.~\ref{fig:pol}).  Nonetheless its steep rise toward
this maximum is itself interesting \cite{TAMM,SpeZal97}.  Since polarization 
isolates the epoch of scattering, we can directly look 
above the horizon scale and test the causal nature of
the perturbations
(see \S \ref{sec:early}).  
Likewise, polarization at even larger scales
can be used to measure the epoch and optical depth during
reionization \cite{HogKaiRes82}
but will require the sub $\mu$K sensitivities
of Planck and future missions.

Finally the ``$E$'' pattern of polarization discussed above
is a special property of density perturbations in the linear
regime.  Its complement (``$B$'' pattern) has the polarization
aligned at 45$^{\circ}$ to the wavevector.  Vector
(vorticity) and tensor (gravity wave) perturbations generate
$B$-polarization as can be seen through the quadrupole moments
they generate ($Y_{2\pm1}$ and $Y_{2\pm 2}$ respectively \cite{TAMM,PolPrimer}).
Measuring the properties of the gravity waves from inflation
through the polarization is our best hope of testing the
particle physics aspects of inflation (see e.g. \cite{KamKos99}).

$B$-polarization is also generated by non-linear effects where
mode coupling alters the relation between the polarization
direction and amplitude. In the context of the simplest inflationary
models, the largest of these is the gravitational lensing
of the primary polarization \cite{Lensing} 
but opacity-modulated secondary
Doppler effects also generate $B$-polarization \cite{Hu00}.

\section{Discussion}

We are already well on on our way to extracting the cosmological
information contained in the primary temperature anisotropies,
specifically the angular diameter distance to recombination,
the baryon density, the matter-radiation ratio at
recombination, and the ``acausal'' (inflationary) 
nature and spectrum of the initial perturbations.  
Even if our simplest inflationary cold dark
matter model is not correct in detail, 
these quantities will be measured in the
next few years by long-duration ballooning, interferometry and
the MAP satellite, {\it if 
the acoustic nature of the peak at $\ell\sim 200$ is 
confirmed by the detection of a second peak}.
In the long term,  the high sensitivity and wide frequency
coverage of the Planck satellite and other future experiments
should allow CMB polarization and secondary anisotropies to 
open new windows on the early universe and large-scale structure.

\vskip 0.2truecm
\begin{quotation}
    \footnotesize\noindent
{\it Acknowledgements:} I would like to thank my collaborators through
the years, especially N. Sugiyama and M. White, the organizers
of RESCEU 1999, and the Yukawa Institute for its hospitality. 
This work was partially supported by NSF-9513835, the Keck
Foundation, and a Sloan Fellowship.
\end{quotation}

\end{document}